\documentclass[conference]{IEEEtran}
\IEEEoverridecommandlockouts

\usepackage{cite}
\usepackage{amsmath, amssymb, amsfonts}
\usepackage{graphicx}
\usepackage{textcomp}
\usepackage{xcolor}
\usepackage{enumerate}
\usepackage{threeparttable}
\usepackage{tabularx} 
\usepackage{float}
\usepackage{caption}
\usepackage{placeins}
\usepackage{latexsym}
\usepackage{subcaption}
\usepackage{wrapfig}
\usepackage{lscape}
\usepackage{pgfgantt}
\usepackage{times}

\usepackage{stfloats}
\usepackage{multirow}
\usepackage{xspace}
\usepackage{calc}
\usepackage{fancyhdr}
\usepackage{varioref}
\usepackage{tikz}
\usepackage{booktabs}
\usepackage{theorem}

\usepackage{algorithm}
\usepackage{algcompatible}
\usepackage{longtable}
\usepackage{verbatim}
\usepackage{algpseudocode}
\usepackage{url}
\usepackage{colortbl}
\usepackage{enumitem}
\usepackage{bm}
\usepackage{makecell}
\usepackage{soul} % Required for \ul command
\usepackage{hhline}
\usepackage{boldline} 
\usepackage{tcolorbox}
\usepackage{mathtools}
\DeclarePairedDelimiter{\ceil}{\lceil}{\rceil}

\renewcommand{\refname}{\hfil References Cited\hfil}    

\newcommand{\algrule}[1][.2pt]{\par\vskip.5\baselineskip\hrule height #1\par\vskip.5\baselineskip}

{\theorembodyfont{\itshape} {\bfseries}{\rmfamily}}
{\theorembodyfont{\rmfamily}
{\bfseries}{\rmfamily}}
{\bfseries}{\rmfamily}
{\bfseries}{\rmfamily}

\newcommand{\specialcell}[2][c]{
\begin{tabular}[#1]{@{}c@{}}#2\end{tabular}}

\newcommand{\romannum}[1]{\uppercase\expandafter{\romannumeral #1\relax}}

\newcommand{\as}{\ensuremath {\leftarrow}{\xspace}}
\newcommand{\sk}{\ensuremath {sk}{\xspace}}

\newcommand{\pk}{\ensuremath {PK}{\xspace}}

\newcommand{\Ra}{\ensuremath \stackrel{\$}{\leftarrow}{\xspace}}

\newcommand{\sgn}{\ensuremath {\texttt{SGN}}{\xspace}}

\newcommand{\sgnsig}{\ensuremath {\texttt{SGN.Sig}}{\xspace}}
\newcommand{\sgnver}{\ensuremath {\texttt{SGN.Ver}}{\xspace}}
\newcommand{\sgnkg}{\ensuremath {\texttt{SGN.Kg}}{\xspace}}

\mathchardef\mhyphen="2D
\newcommand{\eucma}{\mathit{EU\mhyphen CMA}}
\newcommand{\oeucma}{\mathit{OEU\mhyphen CMA}}

 \newtheorem{theorem}{Theorem}
{\theorembodyfont{\rmfamily}
{\theorembodyfont{\rmfamily}
\newtheorem{definition}{Definition}{\bfseries}{\rmfamily}}
{\theorembodyfont{\rmfamily}

\newcommand{\Attacker}{\mathcal{A}}

\newcommand{\hors}{\ensuremath {\texttt{HORS}{\xspace}}}
	\newcommand{\horskg}{\ensuremath {\texttt{HORS.Kg}}{\xspace}}
	\newcommand{\horssig}{\ensuremath {\texttt{HORS.Sig}}{\xspace}}
	\newcommand{\horsver}{\ensuremath {\texttt{HORS.Ver}}{\xspace}}

\newcommand{\pds}{\ensuremath {\texttt{PDS}{\xspace}}}
	\newcommand{\pdsinit}{\ensuremath {\texttt{PDS.Init}}{\xspace}}
	\newcommand{\pdsinsert}{\ensuremath {\texttt{PDS.Insert}}{\xspace}}
	\newcommand{\pdscheck}{\ensuremath {\texttt{PDS.Check}}{\xspace}}

\newcommand{\ohbf}{\ensuremath {\texttt{OHBF}{\xspace}}}

\newcommand{\bfhors}{\ensuremath {\texttt{TVPD-HORS}{\xspace}}}
	\newcommand{\bfhorskg}{\ensuremath {\texttt{TVPD-HORS.Kg}}{\xspace}}
	\newcommand{\bfhorssig}{\ensuremath {\texttt{TVPD-HORS.Sig}}{\xspace}}
	\newcommand{\bfhorsver}{\ensuremath {\texttt{TVPD-HORS.Ver}}{\xspace}}

\newcommand{\rebut}[1]{{#1}}

\begin{document}
% \setlength{\parskip}{3pt}

%\title{Efficient, Scalable and Post-Quantum Authentication for Real-time Next Generation Networks with Bloom Filter}
\title{Fast and Post-Quantum Authentication for Real-time Next Generation Networks with Bloom Filter}
%%%%%%%%%%%%%%%%%%%%%%%%%%%%%%%%%%%%%%%%%%%%%%%
%%%%%%%%%%%%%%%%%%% Authors %%%%%%%%%%%%%%%%%%%
%%%%%%%%%%%%%%%%%%%%%%%%%%%%%%%%%%%%%%%%%%%%%%%

\def\university{University of South Florida}
\def\department{Computer Science and Engineering}
\def\city{Tampa, FL, USA}

\author{\IEEEauthorblockN{Kiarash Sedghighadikolaei}
\IEEEauthorblockA{\textit{\department} \\
\textit{\university}\\
\city \\
kiarashs@usf.edu}
\and
\IEEEauthorblockN{Attila A Yavuz}
\IEEEauthorblockA{\textit{\department} \\
\textit{\university}\\
\city \\
attilaayavuz@usf.edu}
}
\maketitle

\begin{abstract}
Large-scale next-generation networked systems like smart grids and vehicular networks facilitate extensive automation and autonomy through real-time communication of sensitive messages. Digital signatures are vital for such applications since they offer scalable broadcast authentication with non-repudiation. Yet, even conventional secure signatures (e.g., ECDSA, RSA) introduce significant cryptographic delays that can disrupt the safety of such delay-aware systems. With the rise of quantum computers breaking conventional intractability problems, these traditional cryptosystems must be replaced with post-quantum (PQ) secure ones. However, PQ-secure signatures are significantly costlier than their conventional counterparts, vastly exacerbating delay hurdles for real-time applications.   

We propose a new signature called {\em Time Valid Probabilistic Data Structure HORS (\bfhors)} that achieves significantly lower end-to-end delay with a tunable PQ-security for real-time applications. We harness special probabilistic data structures as an efficient one-way function at the heart of our novelty, thereby vastly fastening \hors~as a primitive for NIST PQ cryptography standards. \bfhors~permits tunable and fast processing for varying input sizes via One-hash Bloom Filter, excelling in time valid cases, wherein authentication with shorter security parameters is used for short-lived yet safety-critical messages. We show that \bfhors~verification is 2.7$\times$ and 5$\times$ faster than \hors~in high-security and time valid settings, respectively. \bfhors~key generation is also faster, with a similar signing speed to \hors. Moreover, \bfhors~can increase the speed of \hors~variants over a magnitude of time. These features make \bfhors~an ideal primitive to raise high-speed time valid versions of PQ-safe standards like XMSS and SPHINCS+, paving the way for real-time authentication of next-generation networks.

\end{abstract}


%%%%%%%%%%%%%% Sections %%%%%%%%%%%%%% 
\section{Introduction} \label{sec:intro}
The Internet of Things (IoT) and mobile cyber-physical systems rely on Next-Generation (NextG) networks to enable large-scale autonomy through real-time communication among network entities. Smart-grids~\cite{Yavuz:SmartGrid:2015:Velin:MicroGrid} and the Internet of Vehicles are some examples in which real-time communication is essential to maintain the reliability of the application. For instance, in smart grids, the timely verification of command and control messages and fast yet accurate analysis of measurements (e.g., from smart meters) are vital to avoid cascade failures and damages~\cite{NISTIR-SmartGrid,SmartGridReview}. Similarly, vehicular network standards \cite{IEEEStd-VHCL-WAVE} emphasize the importance of delay awareness for safe operation for autonomous driving\cite{tpsdafoe2023enabling,tpsanderson2023zero}. 

The trustworthiness of such delay-aware applications requires that security-sensitive real-time communication is authenticated and integrity-protected. Digital signatures permit scalable broadcast authentication with non-repudiation and public verifiability and are foundational tools for NextG networked systems. For example, some vehicular communication standards require broadcasting several digital signatures per second (e.g., ECDSA~\cite{ECDSA}) to enable secure vehicular communication \cite{IEEEStd-VHCL-WAVE}. NISTIR 7628 \cite{NISTIR-SmartGrid} recommends digital signatures in smart grids for authentication. However, the overhead of digital signatures may negatively impact the reliability of delay-aware applications \cite{glas2012signalyavuz}. For instance, ECDSA introduces a significant end-to-end delay, impacting functionalities like timely break~\cite{SCRA:Yavuz}, similar concerns applicable for potential cascade failures in smart grids with stringent delay requirements (e.g., a few msecs~\cite{NISTIR-SmartGrid}). 

\raggedbottom

The emergence of quantum computers, capable of breaking conventional secure signatures based on the discrete logarithm and integer factorization problems using Shor's algorithm \cite{Shor-algo}, necessitates the adoption of post-quantum (PQ) safe alternatives. Despite the selection of three schemes—CRYSTALS-Dilithium \cite{ducas2018crystals}, FALCON \cite{fouque2019fast}, and SPHINCS+ \cite{SPHINCSPLUS}—for standardization by NIST in response to this threat \cite{NIST2022selected}, current PQ-safe signatures are significantly more costly than conventional secure ones, exacerbating the risks of cryptographic delays. For instance, SPHINCS+ is several magnitudes and a magnitude slower than ECDSA in signing and verifying, respectively. Similarly, Dilithium and Falcon are also much more computationally costlier than their conventional-secure NIST standards (and even more compared to their faster variants such as Ed25519 \cite{Ed25519}, FourQ \cite{SchnorrQ}). Moreover, as shown in \cite{kannwischer2019pqm4}, Falcon-512 requires 117KB, Dilithium requires 113KB, and SPHINCS+ requires 9KB (excluding its very large signature being around 31 KB) for both code and stack memory on ARM Cortex-M4, which represents a significant overhead for resource-constrained devices.  Hence, delay-aware systems should prioritize post-quantum security solutions that ensure security for short durations while maintaining computational and communication efficiency. We further discuss related work and research challenges in section \ref{sec:relatedwork}.

\subsection{Our Contribution} \label{subsec:ourcontrib} 
We created a new lightweight (one-time) PQ secure signature called {\em Time Valid Probabilistic Data Structure HORS} (\bfhors) to fill the need for a fast and security-adaptable digital signature for delay-aware applications. At the core of our innovation lies harnessing \pds~with special features to attain rapid and tunable OWF, thereby enhancing the performance of \hors. Specifically, we synergies One-hash Bloom Filter (\ohbf) \cite{OneHashingBloomFilter}, which requires only a single size-compatible hash call with a small-constant number of modular arithmetic, with precise parameter tuning. We enhanced our scheme by incorporating weak message resilience. We outline some of \bfhors's desirable properties below:  

$\bullet$~\ul{\em Fast Signature Verification:} \bfhors~offers 2.7$\times$ and up to 5$\times$ faster signature verification for high (up to 128-bit) and time valid security (between 32-64 bit) parameters, respectively, against \hors~with standard-compliant SHA256/512, and with similar performance advantages if size-speed optimized Blake family used in time valid cases. These performance advantages persist against \hors~variants with standard cryptographic hashes, for example, over magnitude speed differences versus HORSIC \cite{HORSIC} and HORSIC+ \cite{lee2021horsic+}.

$\bullet$~\ul{\em Fast Key Generation with Equal Signing Performance:} \bfhors~key generation is 6$\times$ and up to 1.3$\times$ faster than that of \hors~with SHA256/512 for time valid and high-security settings, respectively, also being up to 4$\times$ faster for the time valid case against \hors~with Blake. The signing speed and signature size of \bfhors~are the same as with \hors. Therefore, \bfhors~offers the lowest end-to-end delay among its counterparts thanks to its faster verification and key generation with a similar signing speed.

$\bullet$~\ul{\em A Fast and Tunable PQ-Secure Building Block:} One-time \bfhors~outperforms \hors~in almost all settings but especially in time valid cases. Therefore, it is a suitable candidate to serve as a building block for PQ-secure standards like XMSS and SPHINCS+ that rely on \hors~variants. In particular, \bfhors~is ideal for constructing time valid versions of these PQ-secure standards to support real-time applications via one-time to multiple-time transformations.   

$\bullet$~\ul{\em Full-fledge Implementation:} We fully implemented \bfhors~scheme on a commodity hardware available at: 

\vspace{-2mm}
\begin{center}
\begin{tcolorbox}[colframe=black, colback=white, boxrule=0.1mm, sharp corners, boxsep=1mm, left=0mm, right=0mm, top=0mm, bottom=0mm, halign=center, text width=64mm]
\url{https://github.com/kiarashsedghigh/tvpdhors}
\end{tcolorbox}
\end{center}

\section{Preliminaries and Models} \label{sec:preliminaries}

\textbf{Notations:} $||$ and $|x|$ denote concatenation and the bit length of $x$, respectively. $x \Ra \mathcal{S}$ means $x$ is chosen uniformly at random from the set $\mathcal{S}$. $m \in \{0,1\}^*$ is a finite-length binary message. $\{q_i\}_{i=a}^{b}$ denotes $\{ q_a, q_{a+1}, \ldots, q_b\}$. $\log x$ is $\log_2x $. $[1,n]$ denotes all integer values from $1$ to $n$. $f: \{0,1\}^* \rightarrow \{0,1\}^\kappa$ is an OWF. $H: \{0,1\}^* \rightarrow \{0,1\}^L$ and $h:\,\{0,1\}^* \rightarrow \{0,1\}^{L’}$ denote cryptographic and non-cryptographic hash functions, respectively. 

\vspace{-2mm}
%%% Hash-based digital signature %%%
\begin{definition} \label{def:sgn} A one-time hash-based digital signature $SGN$ consists of three algorithms:
 	\begin{itemize}[leftmargin=8pt]
    \item[-] $\underline{ (\sk, \pk, I_{SGN}) \as \sgnkg( 1^{\kappa}) }$: Given the security parameter $\kappa$, it outputs the private key $\sk$, the public key $\pk$, and the system-wide parameters $I_{SGN}$.  

    \item[-] $\underline{ \sigma \as \sgnsig(\sk, m)  } $: Given the private key $\sk$ and message $m$, it returns the signature $\sigma$. 

    \item[-] $\underline{ b \as \sgnver(\pk, m, \sigma) }$: Given $\pk$, message $m$, and its corresponding signature $\sigma$,  it returns a bit $b$, with $b=1$ meaning valid, and $b=0$ otherwise.
	\end{itemize}
\end{definition}

\vspace{-3mm}
%%% HORS %%%
\begin{definition} \label{def:hors} Hash to Obtain Random Subset (\hors) \cite{HORS} is a hash-based digital signature consists of three algorithms:

	\begin{itemize}[leftmargin=8pt]
    \item[-] $\underline{(\sk, \pk, I_\hors) \as \horskg(1^\kappa)}$: Given the security parameter $\kappa$, it selects $I_\hors \as (t, k, l)$, generates $t$ random $l$-bit strings $\{s_i\}_{i=1}^t$, and computes $v_i \as f(s_i), \forall i=1, \ldots, t$. Finally, it sets $\sk \as \{s_i\}_{i=1}^t$ and $\pk \as \{v_i\}_{i=1}^t$.

    \item[-] $\underline{\sigma \as \horssig(\sk, m)}$: Given \sk~and $m$, it computes $h \as H(m)$ and splits $h$ into $k$ $\log{t}$-sized substrings $\{h_j\}_{j=1}^k$ and interprets them as integers $\{i_j\}_{j=1}^k$. It outputs $\sigma \as \{s_{i_j}\}_{j=1}^k$.

    \item[-] $\underline{b \as \horsver(\pk, m, \sigma)}$: Given \pk, $m$, and $\sigma$, it computes $\{i_j\}_{j=1}^k$ as in $\horssig(.)$. If $v_{i_j}=f(\sigma_j), \forall j=1,\ldots,k$, it returns $b=1$, otherwise $b=0$.
	\end{itemize}
\end{definition}

%%%%%% Probabilistic Data Structure (PDS) %%%%%
\begin{definition} \label{def:pds} A Probabilistic Data Structure (\pds) \cite{StandardBloomFilter} for a set $\mathcal{U}$ comprises at least three algorithms:

\begin{itemize}[leftmargin=8pt]
    \item[-] $(bv, \overrightarrow{h}, I_{\pds}) \as \pdsinit(1^\kappa)$: Given the security parameter $\kappa$, it selects $I_{\pds} \as (n,k)$, creates a zeroed $n$-bit vector $bv[.]$ and samples $k$ hash functions $\overrightarrow{h} \as \{h_i\}^k_{i=1}$. It then outputs $(bv[.], \overrightarrow{h}, I_{\pds})$. 

    \item[-] $\pdsinsert(bv, \overrightarrow{h}, u)$: Given $bv$, $\overrightarrow{h}$, and $u \in \mathcal{U}$, it computes the indices $\{i_j\}^k_{j=1} \as f(\overrightarrow{h}, u)$ and sets $bv[i_j] = 1$ $\forall j=1,\ldots,k$. $f()$ is design dependent.

    \item[-] $b \as \pdscheck(bv, \overrightarrow{h}, u)$: Given $bv$, $\overrightarrow{h}$, and $u$, it computes the indices $\{i_j\}^k_{j=1} \as f(\overrightarrow{h}, u)$ and checks if $bv[i_j] = 1$ $\forall j=1,\ldots,k$. If so, it returns $b=1$, meaning $u$ was probably inserted before, and $b=0$ otherwise.
\end{itemize}
\end{definition}

%%% r-subset-resilient and second pre-image resistant %%%
\begin{definition} \label{def:hashrsrspr} Let $\mathcal{H}=\{H_{i,t,k,L}\}$ be a function family indexed by $i$, where $H_{i,t,k,L}$ maps an arbitrary length input to a $L$-bit subset of $k$ elements from the set $\{0, 1, ..., t-1\}$. $\mathcal{H}$ is $r$-subset (RSR) and second-preimage resistant (SPR), if, for every probabilistic polynomial-time (PPT) adversary $\Attacker$ running in time $\le T$: 

\vspace{-4mm}
\begin{equation*}
\begin{split}
\text{InSec}^{RSR}_{\mathcal{H}}(T) = \underset{\Attacker}{max}\{\text{Pr} & [(M_1, M_2, ..., M_{r+1}) \as {\Attacker}(i, t, k) \\ & \hspace{-27mm} \vspace{-19mm} \text{ s.t.}\; H_{i, t, k, L}(M_{r+1}) \subseteq \bigcup_{j=1}^{r}H_{i, t, k, L}(M_{j})]\} < \text{negl}(t,k)
\end{split}
\end{equation*}

\vspace{-5mm}

\begin{equation*}
\begin{split}
\text{InSec}^{SPR}_{\mathcal{H}}(T) = \underset{\Attacker}{max}\{\text{Pr} & [x \as \{0,1\}^*; x' \as {\Attacker}(x) \text{ s.t. } x \neq x' \\ & \hspace{-10mm} \text{and } H_{i,t,k,L}(x)=H_{i,t,k,L}(x')\} < \text{negl}(L)
\end{split}
\end{equation*}
\end{definition}

%%% PDS Security %%%
\begin{definition} \label{def:pdscrow} Let $\mathcal{PDS} = \{\pds_{i,n,k,L'}\}$ be a function family indexed by $i$, where $\pds_{i,n,k,L'}$ has an $n$-bit vector and $k$ $L'$-bit hash functions. $\mathcal{PDS}$ is collision-resistant (CR) and one-way (OW) if, for every PPT $\Attacker$ running in time $\le T$: 

\vspace{-4mm}
\begin{equation*}
\begin{split}
\hspace{-2mm} \text{InSec}^{CR}_{\mathcal{PDS}}(T) = \underset{\Attacker}{max}\{\text{Pr}[u \as {\Attacker(\pds_{i,n,k,L'})} \text{ s.t. } u \text{ was not } \\ & \hspace{-88mm} \text{inserted before} \text{ and } \text{\pdscheck}(u)=1  \text{}]\} < \hspace{-0mm} \text{negl}(n, k, L') 
\end{split}
\end{equation*}

% \hspace{-3mm} $\text{InSec}^{CR}_{\mathcal{PDS}}(T)$ is also known as the false positive probability.
\hspace{-5mm} The CR property denotes the false positive probability.

\vspace{-2mm}
\begin{equation*}
\begin{split}
\text{InSec}^{OW}_{\mathcal{PDS}}(T) = \underset{\Attacker}{max}\{\text{Pr}[u \as {\Attacker(\pds_{i,n,k,L'})} \text{ s.t. } u \text{ was } \\ & \hspace{-80mm} \text{inserted before} \text{ and } \text{\pdscheck}(u)=1  \text{}]\} < \hspace{-0mm} \text{negl}(n, k, L') 
\end{split}
\end{equation*}
\end{definition}

\noindent \textbf{System Model:} We assume a broadcast environment~\cite{SCRA:Yavuz}, in which the signer sends security-sensitive messages to be authenticated by verifiers. \bfhors~is designed for delay-aware applications in which timely verification is vital, like smart grids, wherein several low-end peripheral devices (e.g., smart meters) periodically upload their telemetry to a cloud-supported state monitoring system for immediate authentication~\cite{Yavuz:SmartGrid:2015:Velin:MicroGrid}. \bfhors~is lightweight and suitable for low-end devices (e.g., 8-bit microcontrollers). However, we assume the verifiers can store several public keys (e.g., a cloud server). 

% \bfhors~is designed for delay-aware applications, in which a timely verification is vital, as in smart grids, wherein several low-end peripheral devices (e.g., smart meters) periodically upload their telemetry to a cloud-supported state monitoring system for immediate authentication~\cite{Yavuz:SmartGrid:2015:Velin:MicroGrid}. 

We consider {\em time valid} delay-aware applications, wherein the security of some transmitted messages remains critical only for a specific time interval\footnote{We indicate time valid forgery attacks targeting temporal (real-time) messages but not the long-term attacks aiming at components like master certificates in public key infrastructures.}(see Section \ref{sec:relatedwork}). Our scheme also preserves its performance advantage in high-security parameters but is especially performant on moderate-level security, making it ideal for time valid applications. 

\noindent \textbf{Threat and Security Model:} Our threat model assumes an adversary $\Attacker$ that can monitor all message-signature pairs and aims to intercept, modify, and forge them. $\Attacker$~is quantum computing capable and aims at forgery in a designated time interval to succeed. The digital signature security model capturing our threat model follows the Existential Unforgeability under Chosen Message Attacks ($\eucma$).

\vspace{-2mm}
%%% EUCMA Definition
\begin{definition} \label{def:onetimeeucma} The one-time $\eucma$ ($\oeucma$) experiment for one-time signature \sgn~is defined as follows:

\begin{itemize}[leftmargin=8pt]
    \item[-] $(\sk, \pk, I_{\sgn}) \,\as\, \sgnkg(1^\kappa)$
    \item[-] $(m^*, \sigma^*) \,\as\, \Attacker^{\sgnsig_{sk}(.)}(\pk, I_{\sgn})$
    \item[-] $\Attacker$ wins the experiment with a maximum of one query allowed if $1 \as \sgnver(\pk, m^*, \sigma^*)$ and $m^*$ was not queried to the signing oracle $\sgnsig_{sk}(.)$.
\end{itemize}

\vspace{1mm}
\hspace{-7mm} $Succ^{\oeucma}_{\sgn}(\Attacker) = Pr[Expt^{\oeucma}(\Attacker) = 1]$
\hspace{-4mm} $InSec^{\oeucma}_{\sgn}(T) = \underset{\Attacker}{max}\{Succ^{\oeucma}_{\sgn}(\Attacker)\}$ $<$ $negl(T)$
\end{definition}

\section{Proposed Scheme}\label{sec:scheme}

%%%%%%%%%%%%%%%%%%%%%%%%%%%%%%%%%%%%%%%
%%%%%%%%%%%%%%% Algorithm
%%%%%%%%%%%%%%%%%%%%%%%%%%%%%%%%%%%%%%%
\begin{algorithm}[b!] %htbp
	\small
	\caption{Time Valid Probabilistic Data Structure HORS}\label{alg:bfhors}
	\hspace{5pt}
 
    %%%%%%%% Keygen %%%%%%%%
	\begin{algorithmic}[1]
		\Statex $ \underline{ (\sk, \pk, I_{\bfhors}) \as \bfhorskg(1^{\kappa}): } $
		\vspace{3pt}
		\State Set $I_{\bfhors} \as (I_{\hors}, n, p, T_\Delta)$
        \State Create zeroed $bv[.]$ having $p$ partitions $\{ P_i \}_{i = 1}^{p}$ each of size $n_i$ % as stated in Definition \ref{def:OHBF} and set all the bits to 0 
        \State Generate $t$ random $l$-bit strings $ \{ s_i \}_{i = 1}^t $:  $s_i \Ra \{0,1\}^l , \forall i \in [1,t]$
        \State Insert  $s_i$ into the $bv[.]$ by setting the bit $(h(s_i || i)\mod n_j)$ of $j^{th}$ partition to 1, $\forall i \in [1,t]$ and $\forall j \in [1, p]$
        \State \textbf{if} time valid setting \textbf{then} set $T_s$ and $T_v$ to $T_0$ \;\;// $T_\Delta$ depends on  $\kappa$ and application needs (e.g., ranging from a minute to days).  
		\State \Return the private key $\sk \as \{ s_i \}_{i = 1}^t$, the public key $\pk \as bv[.]$, and the system-wide parameters $I_{\bfhors}$
	\end{algorithmic}
	\algrule

	%%%%%%%%% Sign %%%%%%%%
	\begin{algorithmic}[1]
		\Statex $\underline{\sigma\as \bfhorssig(\sk, m)}$: $Ctr \as 0$ continue as follows:
		\vspace{-1.5mm}
        %\State $Ctr \as 0$
        \State \textbf{if} {$\kappa$ is high security level \textbf{or} $T_s \in [T_0, T_0 + T_\Delta]$}
        \State \hspace{3.5mm} $ hash \as H ( m || Ctr ) $ 
        \State \hspace{3.5mm} $hash' \as Trunc(hash, k\log t)$ %\;// Truncate the hash to the length $k \log t$ based on the given $\kappa$
		\State \hspace{3.5mm} Split $hash'$ into $k$ substrings  $ \{ hash'_j \}_{j = 1}^{k} $ s.t. $ |hash'_j| = \log t $		
        \State \hspace{3.5mm} Interpret each $hash'_j$ as an integer $i_j$,  $ \forall j  \in [1, k]$
		\State \hspace{3.5mm} \textbf{if} {there are $p,q \in [1, k]$ s.t. $i_p = i_q$ and $p\neq q$} \textbf{then} \text{\hspace{15mm}} $Ctr \as Ctr+1$ and \textbf{goto} step 2 
        \State \hspace{3.5mm} \Return  $\sigma \as (\{ s_{i_j} \}_{j = 1}^k, Ctr) $
	\end{algorithmic}
	\algrule

    %%%%%%%% Verify %%%%%%%%
	\begin{algorithmic}[1]
		\Statex $ \underline{ b \as \bfhorsver( \pk, m, \sigma): } $
		\vspace{3pt}
	%	\State Recall that $ \sigma \as  (\{ s'_i \}_{i = 1}^k, Ctr)$ and $PK \as bv[.]$
        \State \textbf{if} {$\kappa$ is not high security level \textbf{and} $T_v \notin [T_0, T_0 + T_\Delta]$} \textbf{then} \textbf{return} $b=0$
        \State $hash \as H(m || Ctr)$
        \State $hash' \as Trunc(hash, k\log t)$
		\State Split $hash'$ into $k$ substrings  $ \{ hash'_j \}_{j = 1}^{k} $ s.t. $ |hash'_j| = \log t $	
 		\State Interpret each $hash'_j$ as an integer $i_j$, $\forall j \in [1, k]$
        \State \textbf{if} {$\exists$ $p,q \in [1, k]$ s.t. $i_p = i_q$ and $p\neq q$} \textbf{then} \textbf{return} $b=0$ 
        \State \textbf{if} {all the bit indices $(h(s'_j || i_j) \mod n_i)$ of the $i^{th}$ partition in the $bv[.]$ are set, $\forall j \in [1, k ]$, $\forall i \in [1, p]$} \textbf{then} \textbf{return} $b = 1$ \textbf{else} \textbf{return} $b = 0$ 
	\end{algorithmic}
\end{algorithm}

We propose \bfhors~that synergizes special \pds~with efficient hash functions to enable overall high performance with significantly faster operations in time valid settings. Recall that using standard BF with \hors~(e.g., \cite{shafieinejad2017post}) yields highly inefficient results due to excessive hash calls to maintain false positives at par with security parameters. We overcome this challenge by adapting the One-hash Bloom Filter (\ohbf) \cite{OneHashingBloomFilter} that utilizes only one efficient hash operation combined with fast modulo operations suitably selected for the target security levels. Finally, we introduce weak key mitigation~\cite{HORSWeakMessageAttack} into \bfhors~often omitted in its counterparts. We describe \bfhors~in Algorithm \ref{alg:bfhors} and elaborate it as follows:

$\bfhorskg(.)$ sets system-wide parameters $I_{\bfhors}$, comprising \hors~parameters (Definition \ref{def:hors}), \ohbf~parameters $(n,p)$, and time epoch parameter $T_\Delta$ for potential time valid settings (Step 1). Next, a zeroed $n$-bit vector $bv[.]$ with $p$ partitions of each of size $n_i$, satisfying $\Sigma{n_i} \ge n$ and $gcd(n_i, n_j)=1$ for all $i,j \in [1,p]$, is created \cite{OneHashingBloomFilter} (done as $\pdsinit(n, k=1)$) (Step 2). The \hors~private key is generated (Step 3), and each element $s_i$ is inserted, along with its index $i$, into the $bv[.]$ by setting the bit of every partition $j \in [1,p]$ at index $(h(s_i || i) \mod n_j)$ (Step 4). Time synchronization parameters are set if a valid setting is selected (Step 5). This step is skipped for high-security levels (e.g., 72-bit to 128-bit security).

$\bfhorssig(.)$ resembles the \hors~signing. In time-valid settings, signing occurs only in a given time slice $T_\Delta$, while at high-security levels, no such restriction is imposed (Step 1). To eliminate weak message vulnerability, $m$ is concatenated with counter $Ctr$ to ensure its hash contains $k$ distinct $\log t$-sized parts (Steps 2-6). Since only $k\log t$ bits of the hash are used (security reduction from $L$-bit to $k\log t$-bits), the hash output is truncated to $k\log t$ (Step 3). 

$\bfhorsver(.)$ first checks if the verification occurs within its designated time interval (Step 1) in time-valid settings (skipped for high-security levels). Next, it checks weak message conditions with the received $Ctr$ by ensuring $hash'$ contains  $k$ distinct parts (Steps 2-6). Finally, it verifies if the signature elements ($s'_j$) are valid by checking the existence of $(s'_j || i_j)$ in the $bv[.]$ (done as $\pdscheck(.)$) (Step 7). If all the signature elements exist, then the signature is valid.

% \cbstart

\subsection{One-time Key Management for Longevity, Storage, and Scalability}

%Time-bounding the generation and use of \bfhors~public keys may conflict with signing them using long-term secure certificates. However, to extend \bfhors~from OTS to a scalable $N$-time signature, we consider two practical research directions: 
\rebut{
It is necessary to consider the implications of time-bounding the generation and use of \bfhors~public keys while binding them with long-term secure certificates. We consider two practical directions when \bfhors~is extended from OTS to a scalable $N$-time signature:  (i) \(N\) \bfhors~public keys are derived from a seed \cite{SEMECS}, masked with a random pad, and stored on the verifier. During each signing round, the signer derives the signature and pad, while the verifier unmasks the public key using the pad, and verification proceeds as in \bfhors. This method provides long-term protection via certificates, sustainable signature services, and long-term security through masking. (ii) Utilizing secure enclaves (e.g., Intel SGX \cite{Intel:SGX:2016}) as explored in \cite{HASESACMTOMM} can eliminate the need for signers to provide commitments and certificates. This approach may extend OTSs for multiple signatures and faster key generation. In this method, the verifier generates the public key using the master key in a secure enclave and then loads it into main memory for verification, as in \bfhors.
}
% \cbend

\section{Performance Analysis And Comparison} \label{sec:performance}
We first present a comprehensive performance comparison of \bfhors~with \hors~for time valid and full-security settings and then compare \bfhors~with other \hors~variants to showcase its potential. 

\subsection{Evaluation Metrics and Experimental Setup}
We evaluate private/public key and signature sizes, and then key generation (done offline), signature generation, and verification times for all compared schemes. We implemented compared schemes on a desktop with an Intel i9-11900K$@$3.5GHz processor and 64GB of RAM. We used (i) LibTomCrypt\footnote{https://github.com/libtom/libtomcrypt} to implement SHA2(256/512), (ii) Blake2\footnote{https://github.com/BLAKE2/}, and (iii) CityHash\footnote{https://github.com/google/cityhash} and xxHash3\footnote{https://xxhash.com/} as non-cryptographic hash functions.

\subsection{\rebut{Parameter Selection}}
\rebut{Selecting parameters for a specific security level involves considering: (1) the hash function for message hashing, (2) $k\log t$ bits of the hash output, (3) \hors~signature security $k(\log t - \log k)$, (4) \ohbf~hash collision security, and (5) \ohbf~false positive probability. Grover's algorithm \cite{Grover:1996} can reverse a black-box function with input size \(N\) in \(O(\sqrt{N})\) steps and \(O(\log_2 N)\) qubits. We use this model to evaluate the security of our hash function for message signing and the \ohbf~hash function. Thus, a hash function with \(L\)-bit output provides \(\frac{L}{2}\) bits of security against quantum adversaries.
}

\rebut{The false positive probability (fpp) of \ohbf~\cite{OneHashingBloomFilter} is calculated as follows, where $k$ is the number of partitions, $n$ the number of elements inserted, and $m_i$ the $i^{\text{th}}$ partition's size:}
\vspace{-2mm}
\begin{equation}
    fpp = \left(1 - \sqrt[k]{\prod_{i=1}^{k} e^{-\frac{n}{m_i}}}\right)^k
    \label{eq:placeholder}
\end{equation}

\rebut{To adapt this to our signature, we replaced $n$ with $t$, renamed $k$ to $p$, and $m_i$ to $n_i$ to avoid parameter conflicts. We used Algorithm 1 from \cite{OneHashingBloomFilter} to determine the partition size $n_i$, with a Python3 implementation available in our repository \footnote{https://github.com/kiarashsedghigh/tvpdhors/blob/main/misc/ohbf.py}. Using the algorithm, partition sizes are derived by specifying the total \ohbf~size in bits, the number of partitions, and the number of elements. Once partitions are set, the false positive probability can be calculated as given above.}

\rebut{Regarding parameter selection, for instance, to achieve 32-bit security, by selecting $t=64$ and $k=16$ as in TABLE \ref{tab:perfsha2}, the SHA2-256 hash function provides $\frac{256}{2}$-bit security, but only $k\log t = 96$ bits are covered by \hors, reducing the hash security to 48 bits. The \hors~security is $k(\log t - \log k) = 32$ bits. Using xxHash3-64 as the \ohbf~hash function provides 32-bit security. Therefore, the minimum security guarantee is 32 bits, which also requires \ohbf's false positive probability to be 32-bit. Using Algorithm 1 from \cite{OneHashingBloomFilter}, we experimentally, with different sizes and number of partitions, obtained $p=8$ with a 995-byte \ohbf~and $p=6$ with a 1915-byte \ohbf. We selected the first setting to keep public keys smaller, although the second option may be preferred if storage is less of a concern. The parameters yield the following partition sizes in bits, which are further used to insert the keys:}

\begin{center}
    Partitions = [971, 977, 983, 991, 997, 1009, 1013, 1019]
\end{center}

\rebut{Note that \bfhors~includes a built-in partition calculator implemented in C. The complete parameter list can be found at our code repository \footnote{https://github.com/kiarashsedghigh/tvpdhors/blob/main/misc/Parameters.png}.}

% ; the Python3 script is provided for demonstration purposes only

\subsection{Efficiency Evaluation and Comparison}
Our comparison spans various security levels, ranging from 32-bit to 64-bit for time valid applications and 72-bit to 128-bit for non-time-valid applications (medium/high-security). Security levels are adjusted based on the underlying primitives and their parameters (e.g., SHA2-256/512, CityHash-256, xxHash3-64, $p$ in the \ohbf, etc.)

In Tables \ref{subtab:perfsha2}-\ref{subtab:perfblake}, we compare \bfhors~with \hors~when various hash functions are used as $f()$ or $h()$, for varying security levels in time valid ($\kappa$=32, 48, and 64) and high-security ($\kappa$=72, 96, and 128) settings. We first instantiated \hors~with SHA2-256 to show its performance with NIST compliance and then with the Blake family to offer a speed-optimized time valid comparison against \bfhors. We used the xxHash3 and the CityHash families as $h()$ for \bfhors. We outline our findings as follows:

\begin{table*}[t]
\centering
\scriptsize

\caption{Performance comparison of \bfhors~and \hors}
\label{tab:perfsha2}

\begin{subtable}{.5\textwidth}
    \centering
    \caption{\bfhors~vs \hors~($f()$: SHA2 family)}
    \label{subtab:perfsha2}
    
    \begin{tabular}{| >{\centering}p{1.4cm} | >{\centering}p{2.1cm} | >{\centering}p{0.7cm} | >{\centering}p{0.7cm} | >{\centering}p{0.7cm} | c |}\hline

    %%%%%%%%%%%%%%%%%%%%%%%%%%%%%%%%%%%%%%%%%%%%%%%
    %%%%%%%%%%%%%%%%%%%%% T1 %%%%%%%%%%%%%%%%%%%%%%
    %%%%%%%%%%%%%%%%%%%%%%%%%%%%%%%%%%%%%%%%%%%%%%%
    \textbf{Scheme} & \textbf{(t, k, l, p)} & \textbf{\specialcell{PK \\ (KB)}}  &  \textbf{\specialcell{Kg \\ (\bm{$\mu$}s)}} & \textbf{\specialcell{Ver \\ (\bm{$\mu$}s)}} & \bm{$\kappa$} \\ \hline \hline

    \specialcell[]{\hors \\ \bfhors} & (64, 16, 32, 8) & \specialcell[]{2 \\ {0.971}} & 
    \specialcell[]{12.08 \\ {1.99}}  & \specialcell[]{3.11 \\ {0.66}} & 32 \\ \hline           
    \specialcell[]{\hors \\ \bfhors} & (64, 32, 32, 8) & \specialcell[]{2 \\ {0.971}} & \specialcell[]{13.14 \\ {1.96}} &  \specialcell[]{6.32 \\ {1.28}} & 32 \\ \hline          
    \specialcell[]{\hors \\ \bfhors} & (128, 16, 48, 17) & \specialcell[]{4 \\ {1.87}} & \specialcell[]{24.99 \\ {10.45}} &  \specialcell[]{3.15 \\ {0.709}} & 48 \\ \hline           
    \specialcell[]{\hors \\ \bfhors} & (256, 16, 64, 28) & \specialcell[]{8 \\ {3.93}} & \specialcell[]{48.08 \\ {30.84}} &  \specialcell[]{3.13 \\ {0.647}} & 64 \\ \hline           
    \specialcell[]{\hors \\ \bfhors} & (128, 32, 64, 28) & \specialcell[]{4 \\ {1.95}} & \specialcell[]{24.03 \\ {16.23}} &  \specialcell[]{6.19 \\ {1.3}} & 64 \\ \hline 

    \hlineB{3}
    
    \specialcell[]{\hors \\ \bfhors} & (512, 16, 72, 36) & \specialcell[]{16 \\ {7.9}} & \specialcell[]{98.37 \\ {87.96}} &  \specialcell[]{3.17 \\ {1.12}} & 72 \\ \hline   
    \specialcell[]{\hors \\ \bfhors} & (256, 32, 96, 38) & \specialcell[]{8 \\ {5.44}} & \specialcell[]{46.89 \\ {45.81}} &  \specialcell[]{6.01 \\ {2.19}} & 96 \\ \hline 
    \specialcell[]{\hors \\ \bfhors} & (512, 32, 128, 28) & \specialcell[]{16 \\ {40.73}} & \specialcell[]{94.35 \\ {72.48}} &  \specialcell[]{6.01 \\ {2.29}} & 128 \\ \hline   
    \specialcell[]{\hors \\ \bfhors} & (256, 64, \rebut{128}, 30) & \specialcell[]{8 \\ {17.58}} & \specialcell[]{47.26 \\ {37.66}} &  \specialcell[]{11.83 \\ {4.37}} & 128 \\ \hline 

  \end{tabular}
    \begin{tablenotes}[flushleft]
      \item \hspace{-3mm} In all settings, $f()$ of \hors~was SHA2-256 and $h()$ of \bfhors~was xxHash3-64 \\ for 32-bit, xxHash3-128 for (48,64)-bit and CityHash-256 for (72,128)-bit levels.
      \end{tablenotes}
\end{subtable}
\begin{subtable}{.5\textwidth}
    \centering
    \caption{\bfhors~vs \hors~($f()$: Blake2 family)}
    \label{subtab:perfblake}
    
    \begin{tabular}{|>{\centering}p{1.4cm} | >{\centering}p{2.1cm} | >{\centering}p{0.7cm} | >{\centering}p{0.7cm} | >{\centering}p{0.7cm} | c |} \hline
    
    %%%%%%%%%%%%%%%%%%%%%%%%%%%%%%%%%%%%%%%%%%%%%%%
    %%%%%%%%%%%%%%%%%%%%% T2 %%%%%%%%%%%%%%%%%%%%%%
    %%%%%%%%%%%%%%%%%%%%%%%%%%%%%%%%%%%%%%%%%%%%%%%
    \textbf{Scheme} & \textbf{(t, k, l, p)} & \textbf{\specialcell{PK \\ (KB)}}  &  \textbf{\specialcell{Kg \\ (\bm{$\mu$}s)}} & \textbf{\specialcell{Ver \\ (\bm{$\mu$}s)}} & \bm{$\kappa$} \\ \hline \hline
    
    \specialcell[]{\hors \\ \bfhors} & (64, 16, 32, 8) & \specialcell[]{1 \\ {0.971}} & \specialcell[]{8.51 \\ {1.99}}  & \specialcell[]{1.85 \\ {0.66}} & 32 \\ \hline                 
    \specialcell[]{\hors \\ \bfhors} & (64, 32, 32, 8)  & \specialcell[]{1 \\ {0.971}} & \specialcell[]{7.92 \\ {1.96}} &  \specialcell[]{3.69 \\ {1.28}} & 32 \\ \hline 
    \specialcell[]{\hors \\ \bfhors} & (128, 16, 48, 17) & \specialcell[]{2 \\ {1.87}} & \specialcell[]{15.17 \\ {10.45}} &  \specialcell[]{1.81 \\ {0.709}} & 48 \\ \hline 
    \specialcell[]{\hors \\ \bfhors} & (256, 16, 64, 28) & \specialcell[]{4 \\ {3.93}} & \specialcell[]{29.55 \\ {30.84}} &  \specialcell[]{1.85 \\ {0.647}} & 64 \\ \hline 
    \specialcell[]{\hors \\ \bfhors} & (128, 32, 64, 28) & \specialcell[]{2 \\ {1.95}} & \specialcell[]{15.28 \\ {16.23}} &  \specialcell[]{3.65 \\ {1.3}} & 64 \\ \hline 
    
    \hlineB{3}
    
    \specialcell[]{\hors \\ \bfhors} & (512, 16, 72, 30) & \specialcell[]{10 \\ {6.28}} & \specialcell[]{58.02 \\ {76.5}} &  \specialcell[]{1.91 \\ {1.14}} & 72 \\ \hline 
    \specialcell[]{\hors \\ \bfhors} & (256, 32, 96, 32) & \specialcell[]{8 \\ {6.34}} & \specialcell[]{29.02 \\ {40.09}} &  \specialcell[]{3.68 \\ {2.22}} & 96 \\ \hline    
    \specialcell[]{\hors \\ \bfhors} & (512, 32, 128, 28) & \specialcell[]{16 \\ {40.73}} & \specialcell[]{71.52 \\ {72.48}} &  \specialcell[]{4.65 \\ {2.29}} & 128 \\ \hline     
    \specialcell[]{\hors \\ \bfhors} & (256, 64, \rebut{128}, 30) & \specialcell[]{8 \\ {17.58}} & \specialcell[]{36.22 \\ {37.66}} &  \specialcell[]{9.12 \\ {4.37}} & 128 \\ \hline      
  \end{tabular}
  
  \begin{tablenotes}[flushleft]
   \item  \hspace{-2mm} $f()$ of \hors~was Blake2s-128 for (32-64)-bit, Blake2s-160 for 72-bit, and Blake2b-256 for (96-128)-bit. $h()$ of \bfhors~was xxHash3-64 for 32-bit, xxHash3-128 for 48 and 64-bit and CityHash-256 for (72,128)-bit levels.

  \end{tablenotes}

\end{subtable}
  \begin{tablenotes}[flushleft]
    \item \hspace{-2mm} We used SHA-256 for $H()$ in all cases except 128-bit security level with SHA-512. Message size is 256 Bytes in all cases. The private key size is $t\!\cdot\! l$ for both schemes, but it can also be extracted from a constant-size seed. The PK column for \bfhors~shows the size of the \ohbf~(parameter $n$). The signature size is $k \!\cdot\! l$ plus $\log |Ctr|$ variable (negligible) for both schemes. The signing time is also the same since all signing functionalities are identical.   
  \end{tablenotes}

  \vspace{-4mm}
\end{table*}

%%%%%%%%%%%%%%%%%%%%%%%%%%%%%%%%%%%%%%%%%%%%%%%
%%%%%%%%%%%% Other HORS Variants %%%%%%%%%%%%%%
%%%%%%%%%%%%%%%%%%%%%%%%%%%%%%%%%%%%%%%%%%%%%%%
% \vspace{-100mm}
\begin{table*}[!htbp]
    \centering
    \renewcommand{\arraystretch}{1.3} % Default value: 1
    \caption{Performance comparison of \bfhors~and \hors~variants}
    \label{tab:perfothervariants}
    \vspace{-2mm}

\begin{subtable}{\textwidth}
    \centering
    \caption{Analytical (asymptotic) comparison results}
    \label{subtab:othervariantsasym}
    \footnotesize
    
    \resizebox{\textwidth}{!}{
    \Huge    

    \begin{tabular}{| c | c | c | c | c | c | c |} \hline
    \textbf{Scheme} & \textbf{sk Size} & \textbf{PK Size} & \textbf{Signature Size} &  \textbf{Key Generation} &  \textbf{Signature Generation} & \textbf{Signature Verification} \\ \hline \hline
    
    HORS $(t, k ,l)$ \cite{HORS} & $t\!\cdot\! l$ &  $t\!\cdot\!|f()|$ & $k\!\cdot\! l\!+\!\log |Ctr|$ & $t\!\cdot\! f()$ & $k\!\cdot\! O(1)\!+\!\mu\!\cdot\! H()$ & $H()\!+\!k\!\cdot\! f()$  \\ \hline
        
    HORST $(t, k, l)$ \cite{SPHINCSPLUS} & $t\!\cdot\! l\!+\!(t-1)\!\cdot\! |f()|$ & $|f()|$ & $(k+\log t)\!\cdot\! |f()|\!+\!\log |Ctr|$ & $(2t-1) \!\cdot\! f()$ & $k\!\cdot\! O(1)\!+\!\mu\!\cdot\! H()$ & $H()\!+\!k(\log t\!+\!1) \!\cdot\! f()$ \\ \hline 
    
    HORSE $(t, k, l, d)$ \cite{HORSE} & $t\!\cdot\! l\!+\!t\!\cdot\! (d-1)\!\cdot\! |f()|$ & $t\!\cdot\! |f()|$ & $k\!\cdot\! l\!+\!\log |Ctr|$ & $t\!\cdot\! d\!\cdot\! f()$ & $k\!\cdot\! O(1)\!+\!\mu\!\cdot\! H()$ & $H()\!+\! k\!\cdot\! f()$ \\ \hline

    HORS++ $(t, k ,l)$ \cite{HORS++} & $t\!\cdot\! l$ & $t\!\cdot\!|f()|$ & $k\!\cdot\! l\!+\!\log |Ctr|$ & $t\!\cdot\! f()$ & $k\!\cdot\! O(1)\!+\!\mu\!\cdot\! H()$ & $H()\!+\! k\!\cdot\! f()$  \\ \hline
    
    % CTR size?
    HORSIC $(t, k, l, z, w)$ \cite{HORSIC} & $t\!\cdot\! (l\!+\!w\!\cdot\! |f()|)$ & $t\!\cdot\! |f()|$ & $k\!\cdot\! |f()|\!+\!\log |Ctr|$ & $w\!\cdot\! t\!\cdot\! f()$ & $k\!\cdot\! w\!\cdot\! f()\!+\!\mu\!\cdot\! H()\!+\!G()\!+\!C_{k,z}()$ & $H()\!+\!(z+2) \!\cdot\! f()\!+\! G()\!+\!C_{k,z}()$ \\ \hline 
    
    HORSIC+ $(n, t, k, l, z, w)$ \cite{lee2021horsic+} & $w\!\cdot\! l\!+\!t\!\cdot\! (l\!+\!w\!\cdot\! |f()|)$ & $(1+w+t) \!\cdot\! |f()|$ & $k\!\cdot\! |f()|\!+\!\log |Ctr|$ & $w\!\cdot\! t\!\cdot\! f() $ & $k\!\cdot\! w\!\cdot\! f()\!+\!\mu\!\cdot\! H()\!+\!G()\!+\!C_{k,z}()$ & $H()\!+\!k\!\cdot\! w\!\cdot\! f()\!+\!G()\!+\!C_{k,z}()$  \\ \hline

    \rebut{Shafieinejad et al.} $(t, k, l, n, m)$ \cite{shafieinejad2017post} & $t\!\cdot\! l$ & $m$ & $k\!\cdot\! l\!+\!\log |Ctr|$ & $t\!\cdot\!n\!\cdot\! h()$ & $k\!\cdot\! O(1)\!+\!\mu\!\cdot\! H()$ & $H()\!+\!k\!\cdot\!n\!\cdot\! h()$ \\ \hline

    TV-HORS $(t, k ,l, T_{\Delta}, T_{\phi})$ \cite{TVHORS} & $\ceil{\frac{T_{\phi}}{T_{\Delta}}}\!\cdot\! (l\!+\!t\!\cdot\!|f()|)$ & $(l\!+\!t\!\cdot\!|f()|)\!+\!O(1)$ & $k\!\cdot\! |f()|\!+\!l\!+\!\log |Ctr|\!+\!O(1)$ & $\ceil{\frac{T_{\phi}}{T_{\Delta}}}\!\cdot\! t\!\cdot\! f()$ & $k\!\cdot\! O(1)\!+\!\mu\!\cdot\! H()$ & $H()\!+\!k\!\cdot\! f()\!+\!O(1)$  \\ \hline\hline

    \bfhors\;$(t, k, l, n, p)$ & $t\!\cdot\! l$ & $\sum_{i=1}^{p}{n_i}$ & $k\!\cdot\! l+ \log |Ctr|$ & $t\!\cdot\!(h()\!+\!p\!\cdot\! O(1))$ & $k\!\cdot\! O(1)\!+\!\mu\!\cdot\! H()$ & $H()\!+\! k\!\cdot\!(h()\!+\!p\!\cdot\! O(1))$ \\ \hline
    \end{tabular}
    }
\end{subtable}
\begin{tablenotes}[flushleft] \scriptsize{
\item The memory complexity of the private key is the memory expansion caused by deriving the private keys from a constant-sized master key. The average value of the message-dependent $\mu$ is $\frac{t^k}{t(t-1)...(t-k)}$. Hash function $G()$ and the bijective function $C_{k,z}()$ are specific to {HORSIC} and {HORSIC+}. Although HORS variants are different in their design when used as a one-time signature, their parameters may be ineffective, such as $d$ set to 1 in HORSE, and hence, they perform like HORS.
}
\end{tablenotes}

\vspace{-1mm} 
\begin{subtable}{\textwidth}
    \centering
    \caption{\rebut{Experimental performance comparison results ($f()$: SHA2-256, $h()$: xxHash3-(64,128)/CityHash-256)}}
    \label{subtab:othervariantsnum}
    \resizebox{\textwidth}{!}{%
            \Huge
        \setlength{\arrayrulewidth}{1.3pt}
        \begin{tabular}{| c | *{18}{c|}} \hline

        \textbf{Scheme} & \multicolumn{3}{c|}{\textbf{sk Size (KB)}} & \multicolumn{3}{c|}{\textbf{PK Size (KB)}} & \multicolumn{3}{c|}{\textbf{Sig Size (KB)}} & \multicolumn{3}{c|}{\textbf{Key Gen (\bm{$\mu$}s)}} & \multicolumn{3}{c|}{\textbf{Sig Gen (\bm{$\mu$}s)}} & \multicolumn{3}{c|}{\textbf{Sig Ver (\bm{$\mu$}s)}} \\ & $\kappa=32$ & $\kappa=64$ & $\kappa=128$ & $\kappa=32$ & $\kappa=64$ & $\kappa=128$ & $\kappa=32$ & $\kappa=64$ & $\kappa=128$ & $\kappa=32$ & $\kappa=64$ & $\kappa=128$ & $\kappa=32$ & $\kappa=64$ & $\kappa=128$ & $\kappa=32$ & $\kappa=64$ & $\kappa=128$\\ \hline\hline
                
        HORS $(t, k, l)$ \cite{HORS} & 0.25 & 1 & \rebut{4} & 2 & 4 & 8 & 0.06 & 0.25 & 1 & 12.08 & 24.03 & 47.26 & 890.9 & 893.18 & 901.02 & 3.41 & 6.49 & 12.13 \\ \hline 

        HORST $(t, k, l)$ \cite{SPHINCSPLUS} & 2.21 & 4.96 & 11.96 & \multicolumn{3}{c|}{0.31} & 0.68 & 1.21 & 2.25 & 25.82 & 51.53 & 101.83 & 891.12 & 893.98 & 901.42 & 23.17 & 52.03 & 115.08\\ \hline
    
        HORSE $(t, k, l, d)$ \cite{HORSE} & 0.25 & 1 & \rebut{4} & 2 & 4 & 8 & 0.06 & 0.25 & 1 & 12.05 & 23.87 & 48.89 & 890.97 & 893.34 & 900.88 & 3.28 & 6.35 & 11.85 \\ \hline
        
        HORS++ $(t, k, l)$ \cite{HORS++} & 0.25 & 1 & \rebut{4} & 2 & 4 & 8 & 0.06 & 0.25 & 1 & 11.92 & 24.44 & 47.78 & 891.21 & 893.76 & 901.76 & 3.45 & 6.50 & 12.32 \\ \hline
    
        HORSIC $(t, k, l, z, w)$ \cite{HORSIC} & 4.25 & 9 & 20 & 2 & 4 & 8 & 0.34 & 0.56 & 1 & 26.03 & 51.73 & 102.03 & 895.97 & 899.19 & 904.25 & 3.14 & 4.74 & 7.27 \\ \hline
    
        HORSIC+ $(n, t, k, l, z, w)$ \cite{lee2021horsic+} & 4.257 & 9.015 & 20.03 & 2 & 4 & 16 & 0.17 & 0.28 & 1 & 26.15 & 52.12 & 103.66 & 898.54 & 899.97 & 904.49 & 5.01 & 7.42 & 12.51 \\ \hline

        \rebut{Shafieinejad et al.} $(t, k, l, n, m)$ \cite{shafieinejad2017post} & 0.25 & 1 & 4 & 0.968 & 1.97 & 17.57 & 0.06 & 0.25 & 1 & 56.32 & 345.6 & 168.95 & 891.59 & 893.61 & 901.42 & 14.12 & 86.5 & 42.23 \\ \hline

        TV-HORS $(t, k ,l, T_{\Delta}, T_{\phi})$ \cite{TVHORS} & 2.003 & 4.007 & 8.015 & 2.015 & 4.019 & 20.03 & 0.515 & 1.015 & 2.015 & 12.23 & 23.89 & 47.15 & 891.33 & 893.24 & 901.18 & 3.79 & 6.81 & 12.41 \\ \hline\hline

        \bfhors\;$(t, k, l, n, p)$ & \textbf{0.25} & \textbf{1} & \textbf{\rebut{4}} & \textbf{0.97} & \textbf{1.95} & {17.58} & \textbf{0.06} & \textbf{0.25} & \textbf{1} & \textbf{1.96} & \textbf{16.23} & \textbf{37.66} & \textbf{891.33} & \textbf{893.42 }& \textbf{901.54} & \textbf{0.96} & \textbf{1.6} & \textbf{4.67} \\ \hline 
        
    \end{tabular} %%%%%%%%%%%%%%%%%%%%%% Get timing for blake not sha2
    }
        
\end{subtable}
\begin{tablenotes}[flushleft] \scriptsize{
\item Message size is 256 Bytes. For \textbf{HORS}, \textbf{HORST}, \textbf{HORSE}, \textbf{HORS++}, and \textbf{TV-HORS}, the parameters $(t, k, l)$ have been set to (64, 16, 32) for $\kappa=32$, (128, 32, 64) for $\kappa=64$, and (256, 64, 128) for $\kappa=128$. The parameter $d$ of HORSE and the ratio $\ceil{\frac{T_{\phi}}{T_{\Delta}}}$ of TV-HORS have been set to 1 as we are using them as one-time signature. For \textbf{HORSIC} the parameters $(t, k, l ,z, w)$ have been set to (64, 11, 32, 12, 2) for $\kappa=32$, (128, 19, 64, 20, 2) for $\kappa=64$, and (256, 32, 128, 33, 2) for $\kappa=128$. For \textbf{HORSIC+} with $n=256$ the parameters $(t, k, l ,z, w)$ have been set to (64, 11, 32, 12, 2) for $\kappa=32$, (128, 18, 64, 19, 2) for $\kappa=64$, and (256, 32, 128, 33, 2) for $\kappa=128$. \rebut{For \textbf{Shafieinejad et al.} \cite{shafieinejad2017post}, the parameters $(t, k, l, n, m)$ where $n$ and $m$ denote the number of hashes and size of the Bloom Filter, respectively, have been set to (64, 16, 32, 8, 0.96KB) for $\kappa=32$, (128, 32, 64, 27, 1.97KB) for $\kappa=64$ and (256, 64, 128, 30, 17.57KB) for $\kappa=128$. Moreover, xxHash3-64 was used for $\kappa=32$, xxHash3-128 for $\kappa=64$ and CityHash256 for $\kappa=128$.} For \textbf{\bfhors}, the parameters of $(t, k, l, n, p)$ have been set to (64, 16, 32, 8) for $\kappa=32$, (128, 32, 64, 28) for $\kappa=64$, and (256, 64, 128, 30) for $\kappa=128$.}
\vspace{-5mm}
\end{tablenotes}
\end{table*}	

(i) The verification of \bfhors~is 3-5$\times$ and 2.7$\times$ faster than \hors~with standard-compliant SHA-256 (TABLE \ref{subtab:perfsha2}) in time valid and high-security parameters, respectively. \bfhors~is also 2.8$\times$ and 2$\times$ faster than \hors~with speed-size optimized Blake variants in time valid and high-security parameters, respectively (TABLE \ref{subtab:perfblake}). (ii) \bfhors~key generation shows 1.5-6$\times$ improvement for time valid settings and up to 1.3$\times$ for high-security levels with standard-compliant SHA-256 (TABLE \ref{subtab:perfsha2}). Moreover, \bfhors~is  4.2$\times$ faster for 32-bit and 48-bit security levels with comparable performance for other levels with Blake (TABLE \ref{subtab:perfblake}). (iii) The PK size of \bfhors~is smaller than that of \hors~for all security levels except 128-bit, for which we opted for a larger key for faster key generation. Thanks to \ohbf, \bfhors~permits a more flexible PK size trade-off, and we can choose smaller public key sizes with slower key generation (it is offline). (iv) The signing performance is the same for both schemes.

Tables \ref{subtab:othervariantsasym}-\ref{subtab:othervariantsnum} offer a comprehensive performance comparison of \bfhors~with other prominent \hors~variants analytically (asymptotic) and experimentally (estimated with $f()$ as SHA-256 for $\kappa\!=\!32, 64, 128$), respectively: (i) The signature verification of HORSE, HORS++, and TV-HORS resembles that of \hors. However, HORST requires authentication of each private key element $s_i$ using a Merkle tree, making its verification more expensive than others. In HORSIC and HORSIC+, parameters $z$ and $w$ impact the signature verification, respectively. With the provided configurations, \bfhors~exhibits a 3-32$\times$ improvement in time valid settings and a 1.5-24$\times$ improvement in high-security levels. (ii) While HORSE and HORS++ have similar key generation as \hors,  HORST needs a Merkle tree on top of private key elements $s_i$ to derive the public key, doubling the cost. In HORSIC and HORSIC+, the parameter $w$ significantly impacts key generation. Opting for small $w$ with minimal impact on HORSIC and HORSIC+, \bfhors~still delivers a 6-13$\times$ improvement for time valid settings and 1.2-3$\times$ improvement at the 128-bit security level. (iii) HORST, HORSE, HORS++, and TV-HORS exhibit similar signature generation as \hors, while HORSIC and HORSIC+ have costlier signing. All variants consider weak message mitigation with an extra cost explained in TABLE \ref{subtab:othervariantsasym}. \rebut{(iv) Compared to Shafieinejad et al. \cite{shafieinejad2017post} who used standard bloom filter as OWF, \bfhors~shows 20-28$\times$ faster key generation and 14-50$\times$ faster signature verification in time-valid settings and 4$\times$ and 9$\times$, respectively, in 128-bit setting.}

In summary, some notable takeaways are: (i) The verification speed of \bfhors~surpasses that of \hors~with high-security, and with a significantly growing performance advantage in time valid settings. (ii) The key generation of \bfhors~is faster than \hors~in all levels with standard-compliant hash and remains comparable or slightly lesser in size-adjusted speed-optimized hashes for \hors. Note that key generation is mostly done offline, and we can provide size-speed trade-offs to fasten key generation. (iii) The signing performance of \bfhors~is the same as that of \hors. (iv) The performance superiority of \bfhors~over \hors~also remains valid if not grown in various \hors~variants. \rebut{Compared to NIST's PQ-secure standards for a 128-bit security level, \bfhors~signature verification is 103$\times$ faster than SPHINCS+, 15$\times$ faster than Dilithium, and 8$\times$ faster than Falcon. Additionally, for code and stack, Falcon-512 requires 117KB, Dilithium 113KB, and SPHINCS+ 9KB of storage on ARM Cortex-M4, making them impractical for resource-limited devices \cite{kannwischer2019pqm4}.} (v) End-to-end speed advantage, especially due to faster verification, makes \bfhors~suitable for real-time applications and advancing multiple-time signatures such as XMSS and SPHINCS+ in time valid settings. 

% its performance with other \hors~variants demonstrate its potential

\section{Security Analysis} \label{sec:security}

\begin{theorem} 
\bfhors~is $\oeucma$ secure if $H()$ is $r$-subset-resilient and second-preimage resistant, and \ohbf~is collision resistant and one-way:

\vspace{-3mm}
\begin{equation*}
\begin{split}
\text{InSec}^{{\oeucma}}_{\bfhors}(T) = \text{InSec}^{\text{RSR}}_{H}(T) + \text{InSec}^{\text{SPR}}_{H}(T) + \\ & \hspace{-65mm} \text{InSec}^{\text{CR}}_{\ohbf}(T) + \text{InSec}^{\text{OW}}_{\ohbf}(T ) < \text{negl}(t, k, n, p, L, L')
\end{split}
\end{equation*}
\end{theorem}

\noindent \textit{Proof:} Given valid message-signature pair $(m, \sigma)$, there are the below cases leading to a forgery: 

\begin{itemize}[leftmargin=8pt]
    \item[-] \underline{\em $\Attacker$~breaks $r$-subset-resilient of $H$}: $\Attacker$ finds $m^*$ such that $H(m^*)$ has the same $k$ distinct elements as $H(m)$ but $H(m^*)\neq H(m)$. The success probability of $\Attacker$ is $(\frac{k}{t})^k$, which denotes that after $k$ elements determined by $H(m)$, the $k$ elements of $H(m^*)$ are a subset of them and is negligible for appropriate values of $k$ and $t$. Depending on $m$, $H(m)$ may lack $k$ distinct elements, which are called weak messages. They increase the forgery probability as $\Attacker$ may require fewer $s_i$ elements from the $\sk$, which might be found in the current signature $\sigma$. To ensure $H(m)$ possesses $k$ distinct elements, we employ the incremental variable $Ctr$ in concatenation with the message, generating the desired hash as shown in Steps 2-6 of $\bfhorssig(.)$.
    
    \item[-] \underline{\em $\Attacker$~breaks the second-preimage resistance of $H$}: $\Attacker$ can find $m^*$ such that $H(m) = H(m^*)$ and output valid $(m^*, \sigma)$. As $H$ is anf $L$-bit cryptographic hash function, the probability of finding such collision is $\frac{1}{2^\frac{L}{2}}$. In addition, the selection of parameters $k$ and $t$ impacts the security of $H$. As in \hors, the condition $k\log t = L$ must hold. If $k\log t < L$, then the success probability of $\Attacker$ increases from $\frac{1}{2^\frac{L}{2}}$ to $\frac{1}{2^\frac{k\log t}{2}}$. Therefore, the success probability of $\Attacker$ is $max(\frac{1}{2^\frac{L}{2}}, \frac{1}{2^\frac{k\log t}{2}})$. In \bfhors, to ensure $k\log t = L$, we truncate the message's hash to the size $k\log t$ using the $Trunc(.)$ function.

    \item[-] \underline{\em  $\Attacker$~finds collision on \ohbf}: $\Attacker$ generates $\sigma^* \as \{r_i\}_{i=1}^k$, where $r_i$ is a random value, on $m^*$, such that when the verifier checks the existence of signature elements as $\{ r_j || i_j\}_{j=1}^k$ (where $i_j$ is derived as Step 5 in $\bfhorsver(.)$), it outputs 1, indicating that the signature is valid. That is, $\Attacker$ tries to find $r_i$ such that when concatenated with their index from the truncated value of $H(m^* || Ctr)$, step 7 of $\bfhorsver(\cdot)$ returns 1. Given \ohbf~as the underlying $\pds$ with $p$ partitions $\{ P_i \}_{i = 1}^{p}$ each of size $n_i$, the collision probability (false positive probability) of inserting $N$ items is $(1 - \sqrt[p]{\Pi_{i=1}^p e^{-\frac{N}{n_i}}})^{-p}$ \cite{OneHashingBloomFilter}. Moreover, as $h()$ is an $L'$-bit hash function, the collision probability based on the Birthday paradox is $\frac{1}{2^{\frac{L'}{2}}}$. Hence, the advantage of $\Attacker$ is $max(\frac{1}{2^{\frac{L'}{2}}}, (1 - \sqrt[p]{\Pi_{i=1}^p e^{-\frac{t}{n_i}}})^{p})$ given $N=t$.

    \item[-] \underline{\em $\Attacker$~inverts \ohbf}: Given $bv[.]$, $\Attacker$ recovers the secret key elements $\{s_i\}^t_{i=1}$ inserted into the $bv[.]$~as $\{s_i || i\}^t_{i=1}$ during $\bfhorskg(.)$ (with $\sk$, $\Attacker$ can forge signature on any message). Let $Hashes=\{\scalebox{1.2}{$\bigcap$}^{p}_{i=1}\{x \text{ s.t. } bv[x \mod n_i]=1, \forall i \in [1,p]\}\}$ as the set of all possible $h()$'s output $x$ who caused the bits $bv[.]$ to be set. Given that \ohbf~is using $L'$-bit $h()$ and based on the Birthday paradox, the probability of finding $t$ distinct $s_i$ such that $\{s_i || i\}$ hashes to a value in $Hashes$ is $\frac{1}{2^{\frac{L'}{2}}}$. Therefore, the success probability of recovering the whole secret key will be ${(\frac{1}{2^{\frac{L'}{2}}}})^t$. However, $\Attacker$ does not need the entire secret key but only $k$ elements of it such that $\{ s_j || i_j\}_{j=1}^k$ (where $i_j$ is derived as Step 5 in $\bfhorsver(.)$) hashes to a value in $Hashes$. Therefore, the probability is ${(\frac{1}{2^{\frac{L'}{2}}}})^k$. This case closely resembles the previous case where \ohbf~is not collision-resistant as $\Attacker$ can distinguish the new $s_i$ from the actual secret key element with advantage of $\frac{1}{2^{\frac{L'}{2}}}$. Overall, we conclude:
     
\end{itemize}

% \vspace{2mm}
%the $\text{InSec}^{{\oeucma}}_{\bfhors}(T)$ can be written as follows:

\vspace{-4.5mm}
\begin{equation*}
\begin{split}
\text{InSec}^{{\oeucma}}_{\bfhors}(T) =   max(\frac{1}{2^\frac{L}{2}}, \frac{1}{2^\frac{k\log t}{2}}) +  (\frac{k}{t})^k + {(\frac{1}{2^{\frac{L'}{2}}}})^k + \\ & \hspace{-80mm} max(\frac{1}{2^{\frac{L'}{2}}}, (1 - \sqrt[p]{\Pi_{i=1}^p e^{-\frac{t}{n_i}}})^{p}) < \text{negl}(t, k, L, L', n, p)
\end{split}
\end{equation*}

%%%%%%%%%%%%%%%%%%%%%%%%%%%%%%%%%%%%%%%%%%%%%%%%%%%%%%%%%%%%%%%%%%%%%%%%%%%%%%%%%%%%%%%%%%
% https://link.springer.com/chapter/10.1007/978-3-031-40003-2_21
% https://ietresearch.onlinelibrary.wiley.com/doi/pdf/10.1049/ise2.12040
% https://dspace.library.uvic.ca/server/api/core/bitstreams/8d0e9283-3db4-4c2b-b899-bf9a5eb07d6a/content
% https://dl.acm.org/doi/pdf/10.1145/3319535.3363229
% https://eprint.iacr.org/2017/965.pdf

% \\to be Addressed
\vspace{0.8mm}
\section{Related Work and Research Challenges } \label{sec:relatedwork}
Standard signatures based on Elliptic Curve Cryptography (ECC) \cite{ECDSA,Ed25519} are mentioned in smart-grid~\cite{NISTIR-SmartGrid}, 5G, and vehicular standards \cite{IEEEStd-VHCL-WAVE} with an expressed need for faster alternatives. Various high-speed signatures proposed~(e.g., \cite{SCRA:Yavuz}), but they mostly rely on conventional intractability assumptions (e.g., (EC) Discrete Log Problem, factorization), which can be broken by quantum computers~\cite{Shor-algo}.

NIST Post-Quantum Cryptography (NIST-PQC) standard \cite{NIST2022selected}, which includes lattice-based (e.g., Dilithium~\cite{ducas2018crystals}) and hash-based (HB) (SPHINCS+~\cite{SPHINCSPLUS}) alternatives, offer quantum-safe signatures. Despite their merits, these general-purpose signatures are not suitable for delay-aware applications since they are costlier than their conventional-secure counterparts. In particular, HB standards (XMSS \cite{XMSSRFC}, SPHINCS+~\cite{SPHINCSPLUS}) offer high PQ-security without relying on any number-theoretical assumption making them preferable for security-critical applications. Yet, they introduce high signing and verification overhead. Hence, creating delay-aware HB signatures is a valuable research direction, and we aim to do so by enhancing their underlying building blocks and enabling tunable security and performance trade-offs. 

{\em Quest for Efficient One-way Functions (OWF) for Faster HB Signatures}: One of the most efficient HB one-time signature schemes is Hash-to-Obtain Random Subset (\hors) \cite{HORS}, known for its computational efficiency, also serving as the building block for XMSS and SPHINCS+. Several \hors~variants are proposed (e.g.,\cite{lee2021horsic+,HORSIC,HORSE}) relying on various chaining techniques, time valid security, or offering extended functionalities on top \cite{EfficientHashBasedRedactable}. 

The verification and key generation overhead of \hors~is dominated by one-way functions (OWFs) typically implemented with SHA-256/512. While standard cryptographic hashes are efficient for general-purpose applications, they incur substantial computational overhead when invoked in mass, as in XMSS or SPHINCS+. Therefore, an often omitted but crucial means to enhance \hors~like signatures is to identify, improve, and integrate efficient OWFs into their design. 

{\em Time Valid Security and Need for Tunable OWFs}: In a time valid application, security-sensitive yet short-lived messages need authentication and integrity only for relatively short time durations~\cite{TVHORS}. For instance, real-time command and telemetry for rapid decision-making in a smart-grid system permit the adversary only a brief amount of time (e.g., a few seconds) to forge their corresponding signature. Otherwise, the adversary misses her opportunity to influence the system as the target message has already been processed. In such cases, timely verification of the short-lived message is the priority, which raises a time valid design relying on private/public key pairs with shorter security parameters and lifespan.

HB-signatures are ideal for time valid schemes~\cite{TVHORS}, since unlike lattice-based signatures~\cite{ducas2018crystals,ANT:ACSAC:2021} with various mutually dependent parameters, one can only adjust the length of the hash function with a smooth security response. Yet, the lack of efficient OWFs with variable (small) input sizes is an obstacle to building time valid HB schemes. For instance, consider a time valid \hors~that aims security levels  $32 \leq \kappa \leq 72$ depending on the application requirements, where $\kappa$ (bit) denotes the security parameter. For $\kappa=40$, only 40-bit private key elements are inputted to OWF. Regardless of the length of the short input, SHA256 always performs the same amount of computation; in this example, it processes 40 bits with the cost of a full 256 bits. One may try to mitigate the waste via non-standard lightweight cryptographic hash functions like Blake-128; however, they still lack tunable and fast OWF capabilities as they inherently process fixed block sizes. 

In this work, we harness and optimize probabilistic data structures (\pds) to attain better OWF efficiency for \hors. Note that Bloom-filters (BFs) are often used for privacy-enhancing technologies (e.g.,\cite{Yavuz:PrivacyCognitiveRadio:GSNC:Infocomworkshop:2016}), but to a lesser extent for authentication purposes~(e.g., \cite{BFTESLA}). The closest (and to our knowledge only) alternative to our work is in~\cite{shafieinejad2017post}, which suggests using a basic BF in HB signatures. However, we identify that a straightforward use of BF in \hors~does not improve but worsens the performance. This is due to BF's false positive (FP) rate impacting that of \hors. Even FP rates approaching only low-to-moderate security levels (e.g., $\kappa=64$) require an excessive number of BF (non-cryptographic) hash calls, incurring an overhead significantly more than just implementing \hors~with SHA256 as OWF.

{\em There is a significant need for novel signature building blocks that can offer fast and tunable performance to meet the stringent delay requirements of real-time NextG networked systems in the post-quantum era.}

\section{Conclusion}\label{sec:conclusion}
NextG networks rely on real-time communication, but conventional cryptosystems are vulnerable to quantum threats, and PQ-secure signatures increase delays for being resource intensive. We propose a novel signature scheme, {\em Time Valid Probabilistic Data Structure HORS (\bfhors)}, addressing these challenges. By optimizing \hors~with probabilistic data structures, speed-optimized hashes, and weak-message countermeasures, \bfhors~achieves 3-5$\times$ and 2.7$\times$ faster verification in time valid and high-security settings, respectively, compared to \hors~with SHA2-256 and with several-magnitude speed gains against \hors~variants. Moreover, \bfhors~maintains its speed advantage even with lightweight hashes, ensuring low end-to-end delays with comparable signing speeds and key sizes. These results demonstrate \bfhors's potential to fasten HB signature standards for secure NextG networks.

% Next-generation networked systems, like smart grids and vehicular networks, rely on real-time communication to enable automation and autonomy. However, conventional-secure cryptosystems are vulnerable to emerging quantum computers, while existing PQ signatures are significantly costlier than their traditional counterparts, exacerbating delay and reliability hurdles. % In response to these challenges, we propose a novel signature scheme called {em Time Valid Probabilistic Data Structure HORS} (\bfhors), designed to achieve significantly lower end-to-end delays while offering tunable PQ security. We innovate on \hors~by introducing special \pds~with minimal overhead, speed-optimized hashes with fine-grained parameterization, and weak-message countermeasures. \bfhors~achieves 3-5$\times$ and 2.7$\times$ faster verification in time-valid and high-security settings, respectively, compared to \hors~with SHA256. The speed advantages of \bfhors~remain valid against \hors~using lightweight hashes, with identical signing speed and comparable key sizes to ensure low end-to-end delay in all cases. \bfhors~also offers over-magnitude speed gains when used with \hors~variants, showcasing its versatility to support various HB signatures. Our results indicate that \bfhors~has a significant potential to fasten HB signature standards and serve as an ideal building block to secure NextG real-time networks. 

\section{Acknowledgment}
% This research is partially supported by the NSF CNS-2350213 grant.
This work is partially supported by NSF grant CNS-2350213 and Cisco Research Award (220159).

{\bibliographystyle{plain}
{\bibliography{Bibliography/ref,Bibliography/ref2,Bibliography/AttilaYavuz,Bibliography/crypto-etc,Bibliography/kiarash,Bibliography/SaifNouma}

\end{document}